\documentclass{mn2e}
\usepackage{natbib2,natbibmnfix}
\usepackage{astrojournals}
\usepackage{epsfig}
\usepackage{amssymb}
\usepackage{multirow}
\usepackage{multicol}
\usepackage{color}
\usepackage[varg]{txfonts}
\usepackage{url}

\def \msun{\rm M_\odot}

\date{}
\pagerange{\pageref{firstpage}--\pageref{lastpage}} 
\pubyear{2013}

\begin{document}
\label{firstpage}

\title[Disc tearing in black hole state transitions]{A physical model for state transitions in black hole X--ray binaries}
\author[Nixon \& Salvesen]{Chris~Nixon$^{1}$\thanks{chris.nixon@jila.colorado.edu}\thanks{Einstein Fellow} \& Greg~Salvesen$^{1,2}$ \vspace{0.05in}\\ 
$^1$ JILA, University of Colorado \& NIST, Boulder CO 80309-0440, USA\\
$^2$ Department of Astrophysical and Planetary Sciences, University of Colorado, Boulder, CO 80309-0391, USA\vspace{-0.4in}}
\maketitle

\begin{abstract}
We present an accretion cycle which can explain state transitions and other observed phenomena in black hole X--ray binaries. This model is based on the process of disc tearing, where individual rings of gas break off the disc and precess effectively independently. This occurs when the Lense--Thirring effect is stronger than the local disc viscosity. We discuss implications of this model for quasi--periodic oscillations and the disc--jet--corona coupling. We also speculate on applying this model to active galactic nuclei and other accreting systems.
\end{abstract}

\begin{keywords}
accretion, accretion discs; black hole physics; hydrodynamics; X--rays: binaries
\vspace{-0.2in}
\end{keywords}

\section{Introduction}
\label{intro}
Binary systems occur frequently in astrophysics, from supermassive black hole binaries in galaxy mergers to protostellar binaries in star forming regions. When one component of a stellar binary undergoes a supernova explosion, the result may be an X--ray binary, where the companion star feeds the newly formed neutron star or black hole. These systems show evidence for accretion proceeding through a gas disc (e.g. \citealt{PR1972}; \citealt{Pringle1981}). 

X--ray binaries display a wide variety of spectral and timing properties, with many observable features such as jets, disc winds, coronae, quasi--periodic oscillations and strong disc emission and reflection. Black hole state transitions exhibit all of these complex and interacting features, making a theoretical understanding challenging to construct. However, a powerful phenomenological description emerges from model--independent spectral and timing observations (for a recent review, see \citealt{Belloni2010}). Magnetohydrodynamical (MHD) effects are often supposed to be the underlying physics responsible for this behaviour. However, the aim of this paper is to propose a physically motivated and self--consistent mechanism, using mainly hydrodynamical effects, to explain black hole state transitions.

Accretion discs are often thought to be warped, with conclusive evidence from maser discs in active galactic nuclei \citep[AGN;][]{Greenhilletal2003} and suggestive observations in X--ray binaries (e.g. \citealt{Katz1973}; \citealt{WP1999}; \citealt{OD2001}; \citealt{Milleretal2006b}) protostellar discs \citep[e.g.][]{Hughesetal2009} and perhaps ultraluminous X--ray sources \citep[e.g.][]{PS2013}. In X--ray binaries the black hole spin can be strongly misaligned to the binary orbit by a supernova kick during formation \citep[e.g.][]{JN2004}. This misalignment can persist for the entire lifetime of the system as the black hole angular momentum is much larger than the total angular momentum transferred through the disc (e.g. \citealt{Maccarone2002}; \citealt{NK2013}). For discs with angular momentum misaligned to the spin of the black hole, the differential precession induced by the \cite{LT1918} effect is communicated through the disc by a viscosity \citep{BP1975}. This (effective) viscosity most likely arises from MHD turbulence induced by the magnetorotational instability \citep{BH1991}. There is observational evidence for Lense--Thirring precession in X--ray binaries (e.g. \citealt{MH2005}; \citealt{Schnittmanetal2006}). Assuming the disc can efficiently communicate the precession, this results in an aligned inner disc joined to a misaligned outer disc by a smooth warp (see e.g. \citealt{Pringle1992}; \citealt{LP2006}). 

Most investigations into warped discs use simplistic forms of the disc viscosity \citep[e.g.][]{Pringle1992} as this allows for simulation of thin discs with $\alpha > H/R$ \citep{PP1983}. Some simulations of tilted discs explicitly include MHD effects to generate a turbulent viscosity, but these are currently restricted to thick discs (Fragile et al. 2007, 2009; McKinney, Tchekhovskoy \& Blandford 2013; Sorathia, Krolik \& Hawley 2013a). To make progress on thin discs, \cite{NK2012} simulated the disc evolution using the method of \cite{Pringle1992} with the constrained isotropic viscosities derived from the fluid equations for a warped disc by \cite{Ogilvie1999}. Their simulations suggest that often the disc cannot communicate the precession as efficiently as required to maintain a smooth warp -- instead the disc breaks. Follow--up 3D hydrodynamical simulations found that the disc is indeed unable to communicate the precession, and therefore breaks into distinct planes which precess effectively independently -- this is disc tearing \citep{Nixonetal2012b}. These simulations also show significant dynamical evolution of the disc, predominantly infall of gas between interacting torn rings. In this paper, we explore an accretion picture for black hole state transitions which includes the disc tearing process.\vspace{-0.2in}

\section{Accretion picture}
\label{picture}
\begin{figure*}
  \begin{center}
    \includegraphics[angle=0,width=0.64\textwidth]{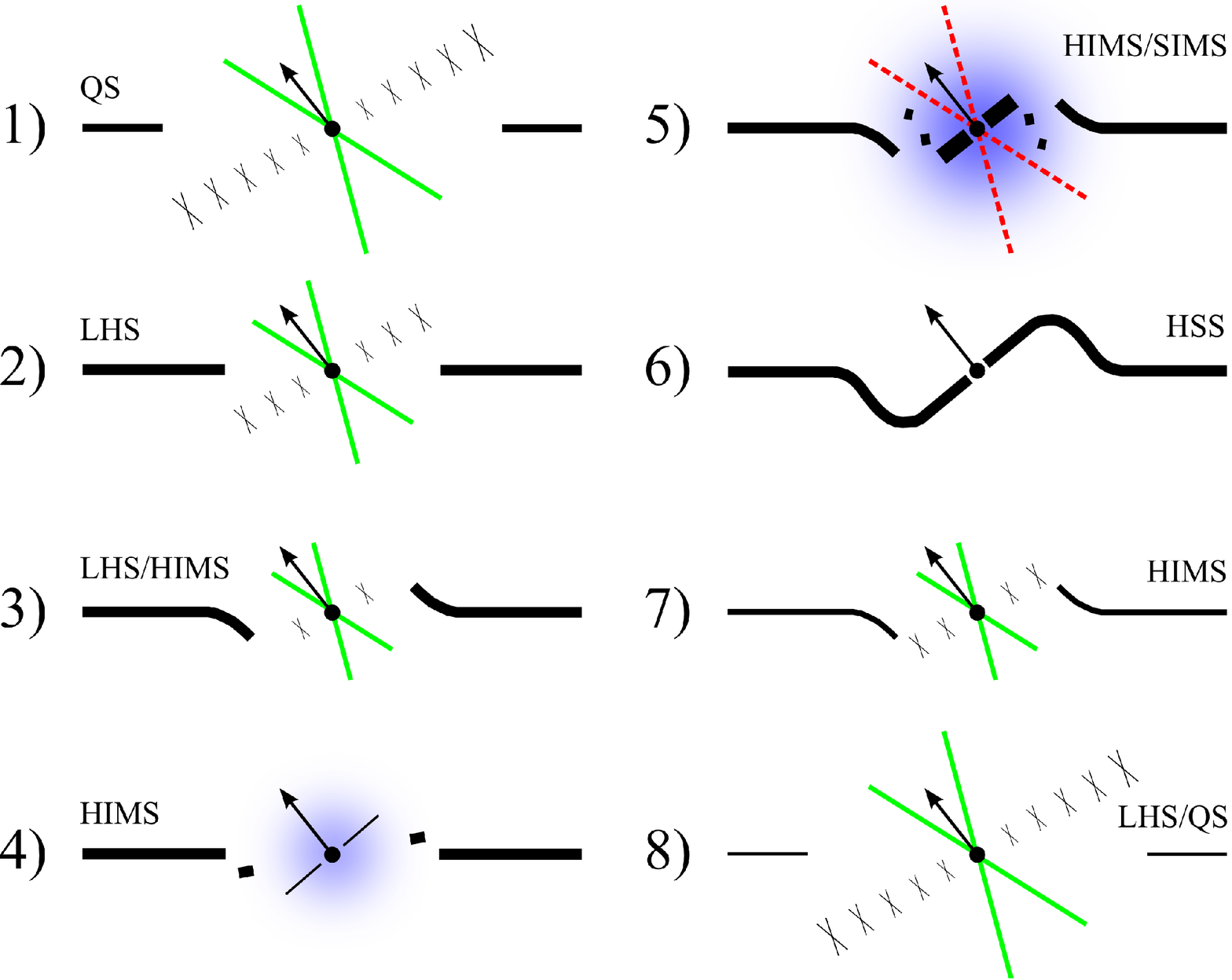}
    \hspace{5.0mm}
    \includegraphics[angle=0,width=0.32\textwidth]{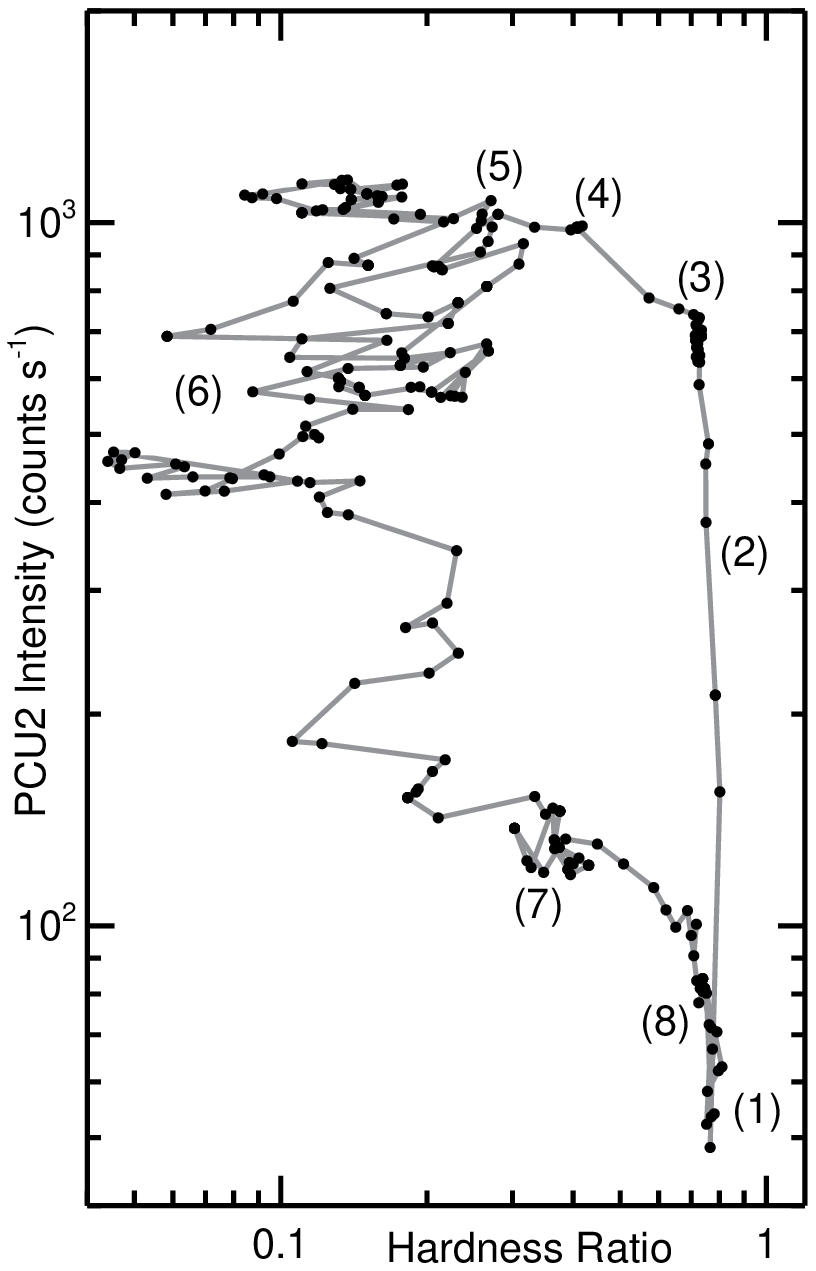}
    \caption{The {\it left panel} shows the different stages of the tearing accretion cycle. In each part the black circle represents the black hole, with the arrow showing the direction of its spin. Solid black lines show the (sometimes warped) Shakura--Sunyaev disc, which in (4) \& (5) has torn, with the thickness approximately denoting its surface density. ``X''s mark regions of hot, low--density gas. The quasi--spherical blue component is the corona. Possible jets are shown by lines leaving the central black hole, green solid lines suggest a steady jet and red dashed lines suggest a transient jet. The figure shows the progression of the system, with each black hole state labelled. The cycle starts with a disc approaching the black hole (1) $\rightarrow$ (2). The initial disc warps (3) and the first ring is torn off (4). The main tearing phase is (5): see Section~\ref{rings}. Once the inner disc aligns, the accretion is more stable (6). As the surface density drops, the hot inner disc is revived (7), and the system returns to its original quiescent state (8) \& (1). We note that a precessing, warped disc is a 3D structure (see e.g. Fig.~\ref{tear}), so the 2D representation shown here is a simple conceptualisation. The {\it right panel} shows where these eight stages lie on a representative hardness--intensity diagram (data from the 2002/2003 GX 339--4 outburst, see e.g. \citealt{Salvesenetal2013}).}
    \label{tearcycle}
  \end{center}
\end{figure*}
We describe an accretion cycle for black hole binary state transitions incorporating disc tearing \citep{Nixonetal2012b}. We begin with plausible, generic initial conditions summarised as follows.\vspace{0.02in}\\
1) {\it A black hole with non--negligible spin.} A modest value of the black hole spin, $a>0.01$, is more than sufficient to drive disc tearing (eq.~8 of \citealt{Nixonetal2012b}). Current estimates of stellar mass black hole spins suggest moderate--to--high values with $0.1 \lesssim a \lesssim 1$ (e.g. \citealt{Milleretal2009}; \citealt{McClintocketal2013}; \citealt{Reynolds2013}).\vspace{0.02in}\\
2) {\it A geometrically thin, optically thick disc of gas at some large radius from the black hole.} This is consistent with the accretion event being triggered by a disc instability, such as the thermal--viscous instability driven by hydrogen ionization (\citealt{MM1982}; \citealt{Smak1982}; \citealt{Lasota2001}).\vspace{0.02in}\\
3) {\it A misalignment between the angular momenta of the accretion disc and the black hole.} Again only a modest disc--spin misalignment angle, $\theta >$ a few degrees, is required \citep{Nixonetal2012b}. Large disc--spin misalignments may be common (e.g. \citealt{Maccarone2002}; \citealt{Martinetal2010}). There is currently little constraint on the disc--spin misalignment. Population synthesis studies \citep[e.g.][]{Fragosetal2010} can generate the likelihood of disc--spin misalignments, but much of the input physics of binary formation and black hole kicks is not understood. Therefore it is likely that the results of Fragos et al., that most binaries have $\theta \lesssim 10^{\circ}$ but up to a third may have $\theta \gtrsim 10^{\circ}$, represents a lower--bound on the possible disc--spin misalignments. Also, analysis of individual sources \citep[e.g.][]{SM2012} can reveal some aspects of the disc--spin misalignment by comparing the binary inclination with the jet inclination. However, this only tests for misalignment along the line--of--sight, offering no constraint perpendicular to this plane. Therefore, one cannot constrain the true disc--spin misalignment, as noted by \cite{SM2012}.

\subsection{Disc evolution}
\label{discevo}
\begin{figure}
  \begin{center}
    \includegraphics[angle=0,width=\columnwidth]{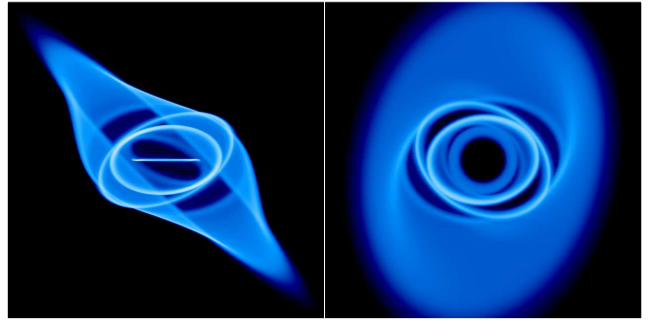}
    \caption{Column density projection showing the disc structure for a tearing disc simulation with a disc--spin misalignment of $45^{\circ}$. The two plots show the same disc from different angles. The {\it left panel} has the black hole spin pointing up the page, while the {\it right panel} has the spin pointing out of the page. The simulation, to be presented in a future paper, follows the method of \cite{Nixonetal2012b}, but uses a smaller accretion radius allowing the innermost disc to be resolved. The innermost disc is closely aligned to the black hole spin because the tilted components of the angular momentum vectors cancel when the precessing rings interact, leading to dynamical infall (see eqns.~3 \& 4 of \citealt{Nixonetal2013}).}
    \label{tear}
  \end{center}
\end{figure}

Fig.~\ref{tearcycle} shows the disc evolution starting from the initial conditions described above. The eight stages in the accretion cycle are:\vspace{0.02in}\\ 
1) An outer Shakura--Sunyaev disc moves inwards towards the black hole, but is still at large radii. Interior to this disc, there is hot, low--density gas present from the last accretion event. This gas may provide the necessary conditions to produce the steady jet sometimes observed in the quiescent state.\vspace{0.02in}\\
2) The outer disc approaches the black hole, with increasing surface density, decreasing the size of the hot, low--density region.\vspace{0.02in}\\
3) The outer disc reaches a radius where the Lense--Thirring precession begins to affect the disc, driving a warp on the inner edge. This occurs at a radius where the inflow time ($R^2/\nu$) becomes comparable to the Lense--Thirring precession time ($1/\Omega_{\rm p}$), which is at
\begin{equation}
R_{\rm tilt} \sim 2^{2/3} \left(\frac{a}{\alpha}\right)^{2/3}\left(\frac{R}{H}\right)^{4/3} R_{\rm g} \approx 1\times10^3 R_{\rm g}
\end{equation}
where $R_{\rm g}=GM/c^2$ is the gravitational radius of a black hole of mass $M$ and we have assumed a \cite{SS1973} viscosity $\nu=\alpha H^2 \Omega$ where $\alpha$ is the dimensionless viscosity parameter, $H$ is the disc semi--thickness at a radius $R$ and $\Omega$ is the disc angular velocity. We adopt typical parameters throughout: $a=0.5$ (e.g. \citealt{Milleretal2009}), $\alpha=0.1$ (e.g. \citealt{Kingetal2007}) and $H/R=0.02$  (e.g. \citealt{SS1973}). Equation~1 is derived through analogous reasoning to \cite{NP1998} who compared the precession time to the vertical viscous time to find the location of a warp in a steady state disc. Here we use the usual planar disc viscosity, appropriate for a disc approaching the black hole from large radius.\vspace{0.02in}\\
4) The first ring of gas is torn from the disc, precessing effectively independently. Any interaction of this ring with the outer disc causes shocks which produce hot gas and dynamical infall after cooling. Any gas that cannot cool quickly enough fuels the corona. There may or may not be an inner disc during this phase. Tearing of disc rings occurs inside \citep{Nixonetal2012b}
\begin{equation}
R_{\rm break} \sim \left(\frac{4}{3}\left|\sin\theta\right|\frac{a}{\alpha}\frac{R}{H}\right)^{2/3} R_{\rm g} \approx 50~R_{\rm g}.
\end{equation}
This radius depends (weakly) on the disc--spin inclination angle. For angles $\lesssim~{\rm a~few}\times H/R$ the disc is unlikely to tear \citep[see e.g. the discussion of ][]{Nixonetal2013}. For a modest inclination angle of $30^{\circ}$ the tearing radius is $R_{\rm break} \approx 30 R_{\rm g}$.\vspace{0.02in}\\
5) The disc reaches a quasi--steady tearing phase, where successive rings are torn off and interact (see Fig.~\ref{tear}). This produces both an aligned inner disc from gas that cools, and a strong corona from gas that remains hot (for more detail see Section~\ref{rings}). The main physics in this interaction is the shocks between adjacent rings. The inner disc, which extends down to the innermost stable circular orbit (ISCO), has a variable mass accretion rate and probably drives short--lived powerful jets while the accretion rate is high. The tearing phase continues as long as misaligned gas is supplied to the tearing region, which persists until the disc aligns from the inner regions outwards. Therefore, the tearing process proceeds for a time $\lesssim t_{\rm align}(R_{\rm tilt})$. The alignment timescale is given by \cite{Kingetal2013}, which (with $M=10~\msun$) gives
\begin{equation}
t_{\rm align}(R_{\rm tilt}) \sim \frac{1}{4\alpha a^2}\left(\frac{H}{R}\right)^2 \left(\frac{R_{\rm tilt}}{R_{\rm g}}\right)^{9/2} \frac{GM}{c^3} \approx 35~{\rm days}.
\end{equation}
We choose the radius for this timescale as $R_{\rm tilt}$, rather than $R_{\rm break}$, as the viscous communication of angular momentum causing alignment is significantly reduced in a strong warp (\citealt{Ogilvie1999}; \citealt{NK2012}; \citealt{LG2013}). To prevent tearing entirely the disc must be unable to feed the tearing region with misaligned angular momentum. This estimate is subject to some uncertainty, and requires further investigation \citep[see e.g. the discussion by][]{Kingetal2013}. As the disc is not in a steady state, it is plausible that alignment out to $\sim R_{\rm tilt}$ is required to halt tearing, but it is also possible that disc tearing can be halted sooner by alignment at a smaller radius, closer to $R_{\rm break}$. However, the two main uncertainties in the alignment timescale are both likely to increase $t_{\rm align}$ by reducing the vertical viscosity responsible for transferring the misaligned angular momentum. First, this viscosity is expected to weaken as the warp amplitude increases \citep{LG2013}. Second, MHD effects are expected to weaken this viscosity further (see e.g. \citealt{Pringle1992}; \citealt{Nixonetal2012b}; \citealt{Sorathiaetal2013} and Section~\ref{discussion}).\vspace{0.02in}\\
6) The inner disc is now closely aligned with the black hole spin beyond the break radius, which inhibits tearing. A strong Shakura--Sunyaev disc component extends down to the ISCO. The outer disc remains misaligned and joined to the inner disc by a warped region. The innermost disc is not subject to the strong depositions of gas present in (5). Therefore the jet probably switches off in this state \citep[cf.][]{Kingetal2004}.\vspace{0.02in}\\
7) As the accretion event runs out of mass supplied from large radius, the disc surface density drops. Eventually, this allows inefficient cooling to revive the inner hot disc.\vspace{0.02in}\\
8) Finally, the accretion event is over and the system returns to the initial state.

\subsection{State transitions}
\label{st}
We now describe the behaviour of observed systems by stepping through the state transition phenomenology, while matching the observations \citep[see e.g.][and references therein]{Belloni2010} to each of the eight stages of the disc tearing accretion picture in Fig.~\ref{tearcycle}.\vspace{0.02in}\\
{\it Quiescent state (QS)}: In quiescence there is no Shakura--Sunyaev disc inside $\sim 10^2 R_{\rm g}$, but a hot gas component may be present interior to this (e.g. \citealt{Esinetal2001}; \citealt{Tomsicketal2009}), possibly driving a steady low--luminosity jet. The accretion cycle starts (1) and ends (8) here.\vspace{0.02in}\\
{\it Low/hard state (LHS)}: As the system comes out of quiescence, a Shakura--Sunyaev disc appears (e.g. \citealt{Milleretal2006}; \citealt{Reisetal2010}). A persistent steady jet is seen \citep[e.g.][]{Fender2001}, suggesting that some inner hot gas is still present (2) \& (3). By the end of the LHS, type--C low--frequency quasi--periodic oscillations (LF QPOs), discussed below, emerge. \vspace{0.02in}\\
{\it Hard--intermediate state (HIMS)}: This state is not marked by any sudden spectral or timing changes and is consistent with a continuation of the LHS. The steady jet turns off and the type--C LF QPOs evolve to higher frequencies in the LHS $\rightarrow$ HIMS transition. An inner disc extending to the ISCO is often observed contemporaneously with the LHS $\rightarrow$ HIMS transition and a corona emerges. In our accretion picture, this is when the first ring is torn from the disc (4). Shocks between this ring and the outer disc create the corona (see Section~\ref{rings}) and possibly a disc that spreads down to the ISCO. This gas sweeps away the hot flow still present in (3), turning off the steady jet. \vspace{0.02in}\\
{\it Soft--intermediate state (SIMS)}: The transition from the HIMS to SIMS is marked by both the appearance of type--B LF QPOs, and relativistic transient (i.e. ballistic) jets. Much weaker type--A LF QPOs are sometimes seen, but type--C LF QPOs no longer persist. The compact corona becomes more prominent and a strong disc extends down to the ISCO, where it remains throughout HIMS $\leftrightarrow$ SIMS transitions \citep{Reisetal2013}. Section~\ref{rings} elaborates on the details of this disc--jet--corona connection. This transition between the HIMS and SIMS is the main tearing phase (5), described in detail in Section~\ref{rings}. The tearing rings quasi--periodically fuel the corona and the inner disc. A powerful jet can be launched when the accretion rate through the inner disc is high. \vspace{0.02in}\\
{\it High/soft state (HSS)}: This state is characterised by a spectrally dominant Shakura--Sunyaev disc component present down to the ISCO. This is well--matched by (6) where a stable, long--lived accretion disc extends down to the ISCO, aligned to the black hole spin. In this phase, the disc is thin, so jets are probably strongly suppressed (e.g. \citealt{Tananbaumetal1972}, \citealt{Fenderetal1999}). However, disc winds are common (e.g. \citealt{Milleretal2008}; \citealt{AKingetal2013}). Tentative evidence for a warped disc in a microquasar was found by \cite{Milleretal2006b} who observed an obscured disc wind consistent with modulation by a disc warp. A standard thin disc closely following the $L_{\rm disc} - T^4_{\rm eff}$ relation is robustly detected across a broad range of mass accretion rates in the HSS, with departures from this relation arising in the intermediate states and LHS \citep[e.g.][]{Dunnetal2011}. However, if a variable spectral hardening factor is permitted, this scaling relation may extend into the intermediate states and LHS (\citealt{Salvesenetal2013}; \citealt{ReynoldsMiller2013}).\vspace{0.02in}\\
{\it Decay HSS $\rightarrow$ HIMS $\rightarrow$ LHS/QS}: In the outburst decay, the system returns to the HIMS when the aligned inner disc is accreted and inefficient cooling allows the hot flow to again pervade the inner regions (7). The SIMS is not re-entered in outburst decay, which is consistent with our accretion picture of the inner disc aligning to the black hole spin (6), halting further disc tearing events that are responsible for the SIMS. The system finally evolves from the HIMS (7) to the LHS (8), accompanied by type-C LF QPOs of decreasing frequency, completing the state transition accretion cycle.\vspace{0.02in}\\
{\it Quasi--periodic oscillations (QPOs)}: There are three main types of LF QPO observed \citep{Casellaetal2005}. Type--C LF QPOs are predominantly observed in both the rise and decay of the system \citep[e.g.][]{Bellonietal2005}, whereas types A \& B are mostly observed in the SIMS and mark the transitions between the HIMS and SIMS \citep{Casellaetal2004}. In our accretion picture, the initial warping and precession of the inwardly propagating disc (3) may manifest itself as the evolving type-C LF QPO, which is consistent with previous suggestions (e.g. \citealt{Ipser1996}; \citealt{Stellaetal1999}; \citealt{Ingrametal2009}). Type--C LF QPOs are also observed, decreasing in frequency, as the system decays back to the LHS \citep[e.g.][]{Kalemcietal2004}.

\cite{Mottaetal2011} suggest that type--C and type--A LF QPOs might be the result of the same underlying physics, at different evolutionary stages, with type--B being physically distinct. In the context of this disc tearing model, this result fits naturally with type--A being precessing torn rings or a precessing oblate corona. This also suggests a plausible reason why type--A are weak and sometimes not seen, as the flux from precessing rings is dominated by the stronger components in the system at that time (e.g. inner disc, corona and jet).\vspace{-0.1in}

\subsection{Disc--jet--corona coupling}
\label{rings}
\begin{figure}
  \begin{center}
    \includegraphics[angle=0,width=0.496\columnwidth]{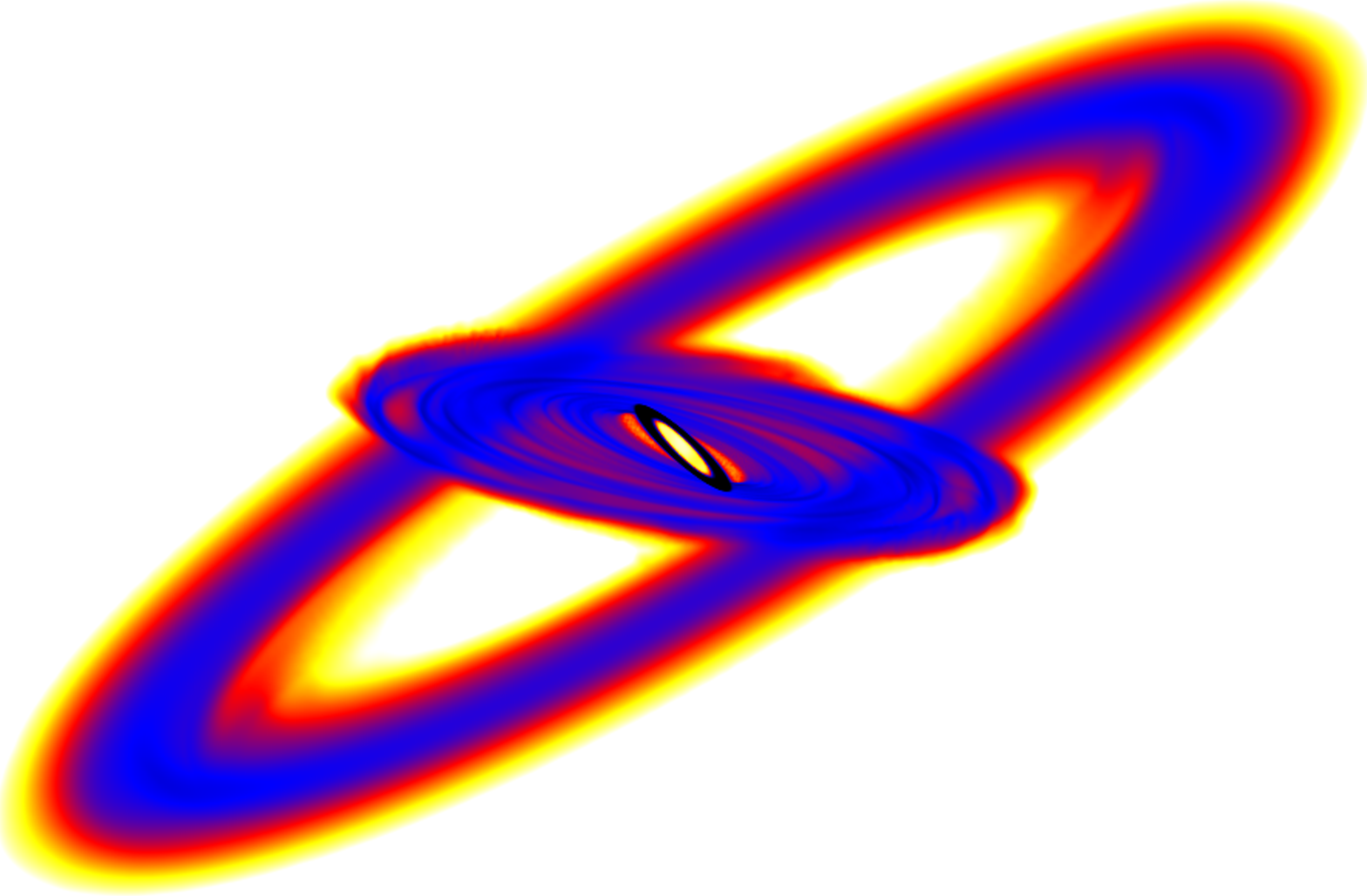}
    \includegraphics[angle=0,width=0.496\columnwidth]{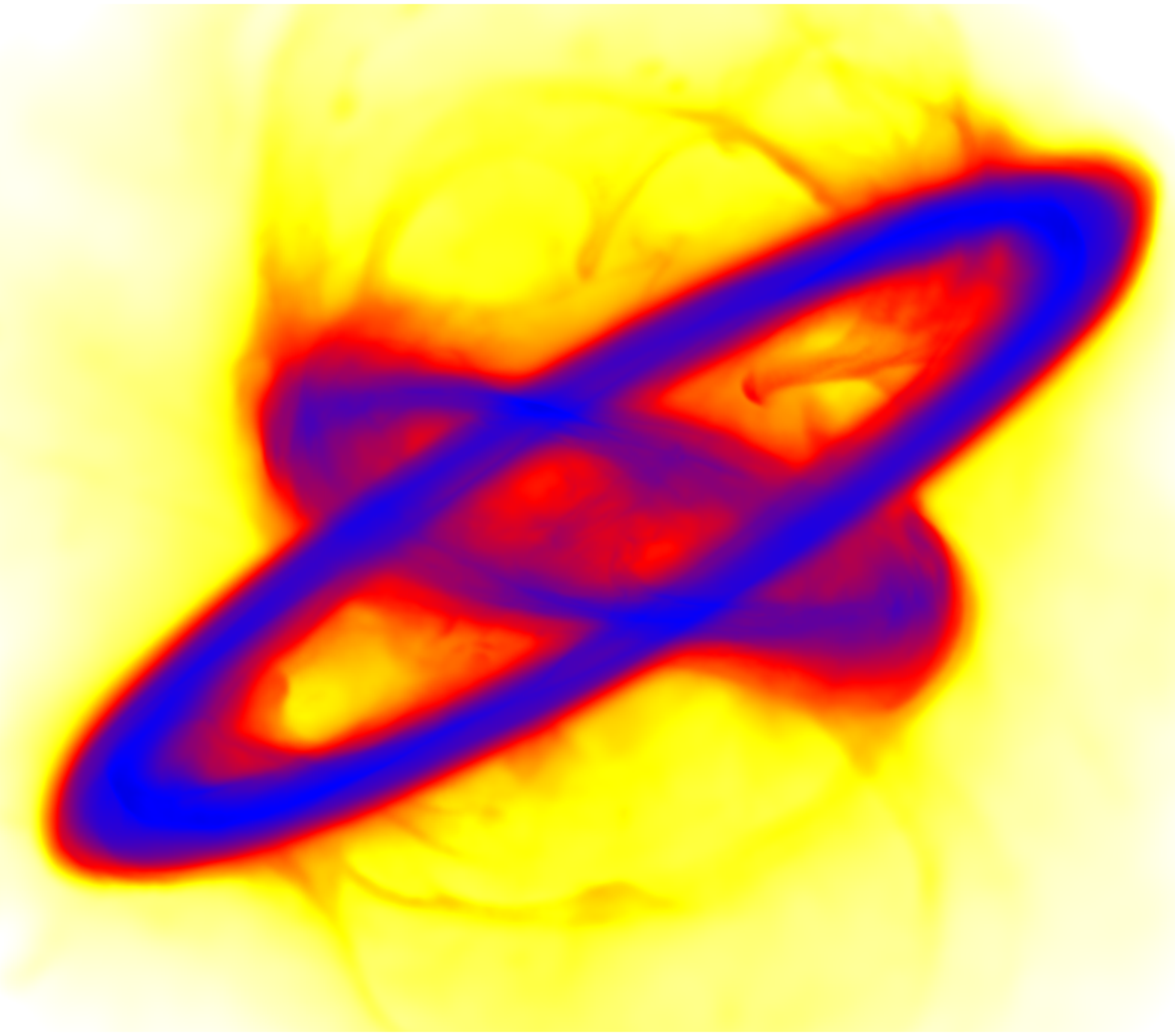}
    \caption{Column density projection showing the disc structure from a misaligned rings simulation in \cite{Nixonetal2012a}. The {\it left panel} uses an isothermal equation of state where gas heating is assumed to be lost as radiation. In contrast, the {\it right panel} retains the shock heating in the gas, causing it to expand vigorously. In reality the equation of state lies somewhere between these two extremes. Most likely some gas is able to cool and subsequently fall to form an inner disc, while the gas that is unable to cool forms a low--density, quasi--spherical structure -- the corona.}
    \label{comp}
  \end{center}
\end{figure}

In the transitions between the HIMS and SIMS, corresponding to (5) in Fig.~\ref{tearcycle} and the torn disc structure in Fig.~\ref{tear}, observations confirm the presence of an inner disc extending down to the ISCO (e.g. \citealt{Dunnetal2011}; \citealt{Reisetal2013}), powerful transient jets \citep[e.g.][]{Fenderetal2004} and a strong dynamic corona \citep[e.g.][]{Reisetal2013}. The production and interaction between the disc, jet and corona naturally arise from our tearing disc picture as follows. Multiple rings break from the disc in the quasi--steady disc tearing phase (see Fig.~\ref{tear}). As these rings precess they shock, cancelling some of their angular momentum but also heating strongly. The cancellation of angular momentum is dependent on the orientation of the rings, with larger disc--spin misalignment angles resulting in more efficient cancellation. Some of the gas cools efficiently to form an inner disc which accretes on to the black hole, while the rest of the gas cannot cool fast enough and so fuels the corona (see {\it right panel} of Fig.~\ref{comp}). While the accretion rate through the inner disc is high, and the disc therefore thick \citep[cf.][]{Kingetal2004}, a powerful jet is launched. Recurring accretion events from interacting rings re--supply the inner disc, driving the observed transient jet. The observed decrease in hardness from the HIMS to the SIMS is therefore given by the cooling gas forming an inner disc, whose spectrum peaks in the soft X--ray band. 

If {\it all} of the shocked gas from tearing rings is unable to cool, which seems unlikely but possible dependent on the local conditions (density etc.), the resultant distribution of gas is quasi--spherical ($H/R \sim 1$) and robbed of a significant part of its angular momentum (implying an effective $\alpha \sim 1$). These are precisely the conditions required to generate radiatively inefficient accretion flows (Narayan \& Yi 1994; Esin, McClintock \& Narayan 1997; Blandford \& Begelman 1999) or magnetically arrested discs (Bisnovatyi-Kogan \& Ruzmaikin 1974; Narayan, Igumenshchev \& Abramowicz 2003; Tchekhovskoy, Narayan \& McKinney 2011)\vspace{-0.2in}.

\section{Discussion \& conclusions}
\label{discussion}
We have presented a new accretion picture for black hole X--ray binaries. The only assumptions needed in this accretion picture are the three generic initial conditions: (1) A black hole with non--negligible spin. (2) A geometrically thin, optically thick accretion disc far from the black hole. (3) A misalignment between the disc and black hole angular momenta. The new ingredient in the accretion cycle is the process of disc tearing (\citealt{Nixonetal2012b}; \citealt{Nixonetal2013}), which is present in the hard--intermediate and soft--intermediate states. Tearing discs produce a variety of behaviour capable of explaining state transitions and many of the other observed black hole X--ray binary phenomena, including QPOs and the disc--jet--corona connection. 

The phenomenological description of accretion in black hole X--ray binaries \citep[e.g.][]{Belloni2010} is powerful in explaining the general observed properties. However, these systems are complex and show different behaviour, not just from source to source, but from outburst to outburst in the same source. Our model may well prove robust in this regard as the equations for the various radii and timescales in Section~\ref{picture} are not weakly (nor too strongly) dependent on the disc parameters. Therefore modest changes in the disc conditions can lead to a rich variety of behaviour. For example, the initial ring torn from the disc (4) may occur while the disc has a low surface--density, so cooling is probably inefficient. In this case, tearing may lead to heating of the disc, which temporarily stabilizes it against tearing while the surface density remains low. These details, and other important physical effects (e.g. radiation warping; \citealt{Pringle1996}), determine the exact evolution of any given system.

In addition to QPOs discussed in Section~\ref{st}, black hole binaries show other timing properties. For example, a positive correlation is observed between rms variability and X--ray flux \citep{UM2001}. The trend of increasing varibility with increasing flux is generally consistent with the accretion picture in Fig~\ref{tearcycle}, but requires further investigation. Strong X--ray flux above $2\,{\rm keV}$ is expected when a strong coronal component is present, which coincides with the most variable phases of accretion.

The viscous evolution of the disc is probably driven by turbulence induced by the magnetorotational instability \citep[MRI;][]{BH1991}. We do not anticipate that including MHD will restrict the tearing behaviour, but this is simply unknown. Progress is being made in this area: investigations using MHD grid codes are starting to look at disc warps (\citealt{Fragileetal2007}; \citealt{Sorathiaetal2013}), and smoothed particle hydrodynamics simulations are beginning to reliably include MHD \citep[e.g.][]{Priceetal2012}. Some investigations into the effective viscosities in a warped disc have already been performed. For example, \cite{Torkelssonetal2000} measured the coefficients of the viscosity tensor from MHD turbulence and concluded that they are consistent with an isotropic viscosity, as used in the disc tearing simulations by \cite{Nixonetal2012b} (cf. \citealt{Ogilvie1999}; \citealt{LP2010}). Also, \cite{Ogilvie2003} developed an analytical model for the dynamical evolution of magnetorotational turbulent stresses in good agreement with \cite{Torkelssonetal2000}. These investigations, numerical {\it and} analytical, allow for the effective viscosity from MHD turbulence to be anisotropic, but conclude the result is (near) isotropic. Recently \cite{Kingetal2013} placed observational constraints on this viscosity, finding that the precession in Her X-1 is consistent with an isotropic viscosity.

\cite{Sorathiaetal2013b} have since claimed the opposite result, that the effective vertical viscosity is much stronger than predicted by an isotropic model. They simulate a misaligned disc subject to Lense--Thirring precession, including MHD to self-consistently generate a turbulent viscosity. However, due to their choice of parameters (thick, low--viscosity disc and a precession which is faster than dynamical at the inner edge of the grid) it is unclear how relevant this work is to the discs discussed here \citep[cf.][]{PP1983}. They report that the dominant mechanism transporting angular momentum is bending waves induced by radial pressure gradients, as shown for $\alpha < H/R$ by \cite{PP1983}; \cite{PL1995} and discussed by \cite{LP2007}. Therefore it appears that this work is consistent with the literature. The main difference is the measurement of a significantly anisotropic effective viscosity, with the vertical viscosity much stronger than isotropy implies. This result appears at odds with previous work on MHD turbulence (\citealt{Torkelssonetal2000}; \citealt{Ogilvie2003}), but as remarked by Sorathia et al., hydrodynamic effects dominate MHD effects and therefore this difference is probably not due to MHD. Indeed for the parameters used in their simulation, the disc may be vulnerable to hydrodynamic instabilities (see e.g. the parametric instability; \citealt{Gammieetal2000}). Such extra dissipation can explain the discrepancy, but without further simulations we can only speculate. However, for typical black hole disc parameters such instabilities are not thought to be important, but certainly worthy of investigation \citep[e.g.][]{OL2013}. For now it appears prudent to continue with the assumption of a (near) isotropic viscosity supported by numerical, analytical and observational results (\citealt{Torkelssonetal2000}; \citealt{Ogilvie2003}; \citealt{Kingetal2013}).

The accretion picture proposed here is not particular to black hole X--ray binaries. We therefore briefly discuss some other systems in the context of this accretion picture:\vspace{0.02in}\\
{\it Persistent sources}: It is plausible that tearing is entirely suppressed for some system parameters. For example, the mass transfer rate in the X--ray binary SS433 is probably highly super--Eddington, so the disc remains thick all the way down to the ISCO \citep{Begelmanetal2006} and is therefore unable to experience tearing. Other sources which show persistent or ``failed'' state transition behaviour may have negligible black hole spin or disc--spin misalignment, thus inhibiting tearing.\vspace{0.02in}\\
{\it Other stellar binaries}: There appears no reason to assume that accretion far from the central object occurs differently for different central objects \citep[e.g.][]{Lasota2001}. Therefore, any model should reduce to the standard accretion model \citep[e.g.][]{Franketal2002} for white dwarf and magnetic neutron star accretion. For neutron stars with weak magnetic fields, a similar cycle as proposed above may arise, as the Lense--Thirring precession is also expected to act in these systems.\vspace{0.02in}\\
{\it Ultraluminous X--ray sources (ULXs):} These may be extreme stellar mass black hole binaries with the largest disc--spin misalignments, so the strong shocks induced by disc tearing produce both powerful X--ray emission and significantly enhance the accretion rate. If ULXs contain intermediate mass black holes they may correspond to (5) in Fig.~\ref{tearcycle}, evolving on much longer timescales.\vspace{0.02in}\\
{\it Active galactic nuclei}: The timescales predicted for state changes in AGN are substantially longer than for X--ray binaries. Therefore, state transitions can only be understood by considering AGN populations. However, it is possible that the different stages of the proposed accretion cycle correspond to different types of AGN. For example, the type I {\it versus} type II AGN dichotomy may be states (5) \& (6) in Fig.~\ref{tearcycle} viewed from the required angle to reveal or obscure the central accretion \citep[cf.][]{Nayakshin2005}.

There are significant observational data with which we can test this model; high--quality spectral and timing information is available for numerous systems. Substantial progress can also be made on the theoretical side. For example, simulations of interacting rings of gas, with a realistic equation of state, are required to confirm our reasoning in Section~\ref{rings}. Also, the shocks between rings and the geometry of the disc may provide a physical motivation for disc reflection modelling. We are actively exploring both observational and theoretical considerations of the accretion picture presented in this paper to put this physically motivated black hole state transition model to the test.

\section*{Acknowledgments}
We thank Andrew King, Jon Miller \& Jim Pringle for valuable discussions. We also thank Phil Armitage and Mitch Begelman for thoughtful comments on the manuscript. We thank the referee for a thorough examination of our paper, which stimulated us to make it (we believe) considerably clearer. CN thanks NASA for support through the Einstein Fellowship Program, grant PF2-130098. GS thanks the National Science Foundation for support through the Graduate Research Fellowship Program. Some of the figures used {\sc splash} \citep{Price2007} for the visualization.\vspace{-0.2in}

\bibliographystyle{mn2e} 
\bibliography{nixon}

\label{lastpage}
\end{document}